\documentclass[%
 twocolumn,
 reprint,
 superscriptaddress,
showpacs,
 amsmath,amssymb,
 aps,
prb,
]{revtex4-1}
\usepackage{color}
\usepackage{ulem}
\usepackage{graphicx}
\usepackage{dcolumn}
\usepackage{bm}

\def\rnum#1{\expandafter{%
\romannumeral #1}}
\def\Rnum#1{\uppercase\expandafter{%
\romannumeral #1}}

\newcommand{\bol}[1]{\boldsymbol #1}

\begin{document}

\preprint{APS/123-QED\author{Takafumi Suzuki}}

\title{Gapless edge states and their stability 
in two-dimensional quantum magnets}

\author{Takafumi Suzuki}
\affiliation{Research Center for Nano-Micro Structure Science and Engineering, 
Graduate School of Engineering, University of Hyogo, Himeji, 
Hyogo 671-2280, Japan}
\author{Masahiro Sato}%
\affiliation{Department of Physics and Mathematics, Aoyama Gakuin University, 
Sagamihara, Kanagawa 229-8558, Japan}%

\date{\today}

\begin{abstract}
We study the nature of edge states in extrinsically and spontaneously dimerized states of two-dimensional spin-$\frac{1}{2}$ antiferromagnets, by performing 
quantum Monte Carlo simulation. We show that a gapless edge mode emerges 
in the wide region of the dimerized phases, and the critical exponent of 
spin correlators along the edge deviates from the value of Tomonaga-Luttinger 
(TL) liquid universality in large but finite systems at low temperatures. 
We also demonstrate that the gapless nature at edges is stable against several 
perturbations such as external magnetic field, easy-plane XXZ anisotropy, Dzyaloshinskii-Moriya interaction, and further-neighbor exchange interactions. 
The edge states exhibit non TL-liquid behavior, depending strongly on model 
parameters and kinds of perturbations. 
Possible ways of detecting these edge states 
are discussed. Properties of edge states we show in this paper could also 
be used as reference points to study other edge states of more exotic 
gapped magnetic phases such as spin liquids. 
\end{abstract}

\maketitle

%
\section{Introduction}\label{Intro}
In recent years, gapful ground states without any {\it local} order parameter and their boundary properties have been vividly studied in quantum many-body physics from both theoretical and experimental viewpoints. 
Among such disordered states, two-dimensional (2D) and three-dimensional (3D) topological insulators (TIs), for instance,  have attracted much attention as novel many-body phases in solids.~\cite{review1,review2} Their fundamental properties are that the bulk has a finite excitation gap, but its boundary (surface or edge) is metallic (i.e., gapless) and around the boundary, up-spin electrons move antiparallel to the motion of down-spin electrons (this nature is called helical). 
This gapless boundary state is quite stable against any perturbation with 
time-reversal symmetry, and the existence of a helical edge mode is recorded 
in a $\mathbb{Z}_2$ topological invariant defined on the bulk 
(bulk-edge correspondence).

In quantum spin systems, the Haldane-gap state~\cite{Haldane1} is also famous as a state without any local order. This state is defined as the ground state of one-dimensional (1D) spin-1 antiferromagnetic (AF) chains and is actually realized in some quasi-1D magnets.~\cite{Renard,Hagiwara} 
Its characteristic features can be captured by a valence-bond solid (VBS) picture.~\cite{Affleck1} Namely, the Haldane state can be approximated by the uniform tensor product state of local singlet dimers composed of two fictitious $S=\frac{1}{2}$ spins on neighboring sites, which are generated via the decomposition of original $S=1$ spin on each site. 
From the solid singlet distribution, we can easily understand the existence of a finite excitation gap (called the Haldane gap) on the Haldane state. 
Similarly, the uniform alignment of singlets indicates the absence of any local order parameter, but we can construct a {\it non-local} string order parameter~\cite{Nijs} to distinguish the Haldane state from the other paramagnetic 
phases. The VBS picture also shows that an almost free $S=\frac{1}{2}$ spin 
appears at the edge of the finite-size spin-$1$ Haldane state under free boundary 
condition.~\cite{Hagiwara} 
In addition to these results, recently new ways of characterizing the Haldane states have been actively discussed based on symmetries and artificial quantities such as entanglement spectra.~\cite{Oshikawa1,Wen1,Katsura}

All of the gapped, topological phases in free fermion systems including TI have been successfully classified theoretically.~\cite{Furusaki,Kitaev,Ryu} 
On the other hand, topological phases and boundary states in quantum spin systems have been less understood, except for a few VBS (such as the Haldane state) and short-range valence-bond states in 1D spin systems including spin ladders. 
Therefore, the understanding of topological nature and boundary properties in quantum spin systems, especially, in {\it higher-dimensional} magnets, is an important, fundamental issue in magnetism. 
One might image VBS~\cite{Tasaki} or exotic spin-liquid states as typical 
examples of 2D or 3D gapped non-magnetic spin states with a gapless edge mode. 
It is, however, difficult to prepare such states in nature because the 
corresponding Hamiltonians contain various tuned coupling constants. 
In addition, usual magnetic ordered states such as N\'eel and spiral 
ordered states are not suitable since both edge and bulk are trivially 
gapless due to the Nambu-Goldstone mode. Instead of these states, 
simple, realistic systems with both a gapless edge mode and a bulk gap 
would be suitable for starting to understand edge modes 
in 2D and 3D spin systems. 
In this paper, we thus study the nature of edge modes in 2D spin-Peierls (dimerized) states, by using the quantum Monte Carlo (QMC) method based on the worm algorithm.~\cite{QMC1,QMC2,QMC3}

There is a similarity between TIs and the Haldane state: A finite bulk gap, gapless boundary states, and the $\mathbb{Z}_2$ topological invariant of TI seem to correspond to a Haldane gap, free edge spins, and the string order parameter of the Haldane state, respectively. 
We will hence consider how the edge modes of dimerized states are different from and similar to those of TIs and the Haldane state. 
It would be impossible to define any topological order parameter for 2D dimerized states, and in that sense, the gapless edge states in dimerized states (even if they exist) are naively expected to be less stable compared to those of TIs. 
We should, however, note that it is generally hard to predict whether or not there exist gapless edge states and how stable they are, since quantum spin systems we consider below are strongly correlated systems that are different from free fermion models for TIs.

The rest of this paper is organized as follows. 
In Sec.~\ref{Model}, we explain two kinds of 2D quantum AF models 
with an extrinsically or intrinsically dimerized phase. 
Both models can be analyzed by using QMC techniques. 
In Sec.~\ref{TLLiquid}, we briefly summarize characteristic features of 
correlation functions of the standard Tomonaga-Luttinger (TL) liquid phase 
in purely 1D critical AF spin systems before the analysis of 
two spin-Peierls models. The properties of TL liquid are useful for discussing 
the edge states in the dimerized phase. 
Section.~\ref{Numerical} is devoted to our numerical analysis and is 
the main content of this paper. We show in Sec.~\ref{Gapless} 
that a gapless edge mode really exists in the dimer phases of both models, 
and the critical exponent of the spin-correlation function along the edge 
moves away from the value of standard TL liquid 
in the weakly dimerized region. 
We then discuss the stability of the gapless edge mode against realistic 
perturbations in Sec.~\ref{Stability}. We demonstrate that 
the gapless nature survives after introducing several perturbations 
with different symmetries, while the edge critical exponents drastically 
change depending on the kinds of perturbations. 
We briefly consider some experimental ways of observing gapless edge 
modes in Sec.\ref{How}. Finally, we summarize our numerical results and 
predictions in Sec.~\ref{Conclusion}.

\section{Model}\label{Model}
\begin{figure}[t]
\includegraphics[width=7cm, trim=5 40 150 0]{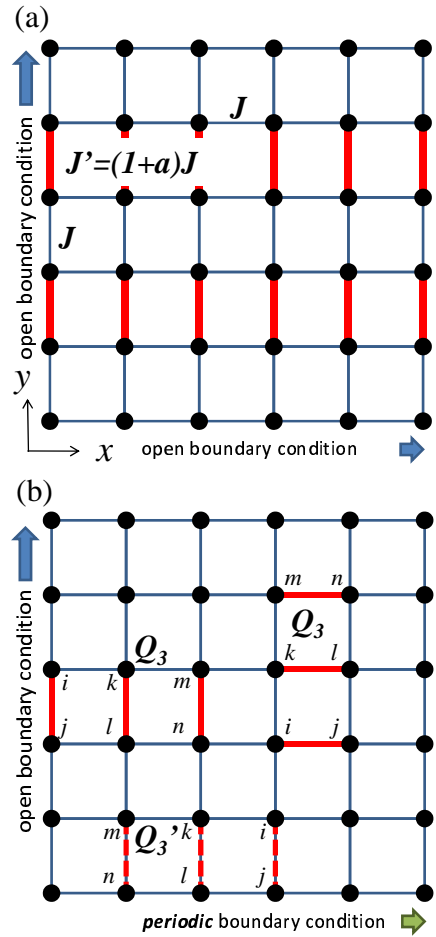}
\caption{(color online) 
(a) 2D dimerized model with bond alternation along the $y$ direction. 
(b) 2D $JQ_3$ model with six-spin interaction term. 
The modified coupling constant $Q_3'$ on edges is explained 
in Sec.~\ref{Model}.}
\label{fig_model}
\end{figure}

Dimerized phases are roughly classified into an extrinsically dimerized phase without any spontaneous symmetry breaking (SSB) and an intrinsically dimerized one with a spontaneous translational symmetry breaking. 
To study those two states we utilize two SU(2)-symmetric spin-$\frac{1}{2}$ AF models on a square lattice defined in the $x$-$y$ plane: 
the dimerized model~\cite{dimer_model1,dimer_model2} and 
the $JQ_3$ model.~\cite{JQ3_model} 
Their Hamiltonians are given as \begin{subequations}
\label{eq:2models}
\begin{eqnarray}
\label{eq:dimer}
{\mathcal H}_{\rm dim} & = & \sum_{\langle i,j \rangle}
J{\bol S}_{i}\cdot{\bol S}_{j}
+\sum_{j}J\alpha\delta_{j_y,{\rm even}}
{\bol S}_{j}\cdot{\bol S}_{j+{\bol e}_y},\hspace{0.5cm}\\
{\mathcal H}_{JQ3} & = & \sum_{\langle i,j \rangle} 
J{\bol S}_{i}\cdot{\bol S}_{j} 
 + \sum_{ \langle i, j, k, l, m, n \rangle } 
Q_{3 ijklmn}  C_{ij}C_{kl}C_{mn},\hspace{0.5cm}
\label{eq:JQ3}
\end{eqnarray}
\end{subequations}
where $\bol S_{j}$ is the spin-$\frac{1}{2}$ operator on site $j=(j_x,j_y)$ ($j_{x,y}\in \mathbb{Z}$), ${\bol e}_{x[y]}=(1,0)$ $[(0,1)]$ is the unit vector for the $x$ $[y]$ direction, $J>0$ is the AF exchange coupling constant between neighboring spins, and $C_{ij}=1/4-{\bol S}_{i}\cdot{\bol S}_{j}$. 
In the dimerized model (\ref{eq:dimer}), $\alpha$ denotes the magnitude of external dimerization along the $y$ direction as shown in Fig.~\ref{fig_model}(a), 
in which the dimerized bond strength $J'=J(1+\alpha)$. 
If $\alpha$ is strong enough and the open boundary condition for $y$ direction is imposed as in Fig.~\ref{fig_model}, an effective spin chain is expected to appear along the $x$ direction at the edge thanks to the formation of dimerization on all strong bonds $J'$. 
In the $JQ_3$ model (\ref{eq:JQ3}), the second $Q_3$ term includes six-spin interactions, where six sites $(i,j,k,l,m,n)$ are defined 
on two neighboring plaquettes (rectangle) shown in Fig.~\ref{fig_model}(b). 
The symbol $\sum_{\langle i,j,k,l,m,n\rangle}$ stands for the summation over all rectangles on the square lattice.

The ground-state phase diagrams for the dimerized and $JQ_3$ models have been investigated by QMC calculations, and both models show the N\'eel-dimer quantum phase transition.~\cite{Tanaka} 
For the dimerized model, the critical point is located at $J'=J'_c\simeq 1.91J$~\cite{dimer_model1,dimer_model2} and singlet dimers appear on all the bonds $J'$ for $J'>J'_c$. 
The transition of the $JQ_3$ model takes place at $Q_3={Q_3}_c\simeq 1.5J$ and the spins spontaneously form a columnar dimer state along the $x$ or $y$ direction when $Q_3>{Q_3}_c$.~\cite{JQ3_model} 
It is worth noticing that the dimer phase of Eq.~(\ref{eq:dimer}) does not accompany any SSB similarly to TIs, while in the case of Eq.~(\ref{eq:JQ3}), the translational symmetry is spontaneously broken.

The $JQ_3$ model seems to be a toy model, but it is one of the few spin models with a spontaneously dimerized phase and can be accurately analyzed by QMC simulation without a negative sign problem. Furthermore, its dimerized ground-state wave function is expected to be close enough to that of real dimerized magnets. 
To study possible gapless edge modes of the model~(\ref{eq:dimer}), we impose an open boundary condition for the $y$ direction as shown 
in Fig.~\ref{fig_model}(a). 
On the other hand, we have confirmed from the QMC simulation that 
in the dimer phase of the $JQ_3$ model, singlet dimers tend to reside 
on edges when we set the open boundary condition to make the edges. 
For this dimerization pattern, no gapless edge state is expected. 
In order to remove the dimers on edges and make the same dimerization 
pattern as that of the dimerized model~(\ref{eq:dimer}), 
we modify the value of $Q_3$ to $Q_3'(<Q_3)$ at the edge as depicted in 
Fig.~\ref{fig_model}(b). 
From QMC results of $Q_3'/Q_3=0.0$, $0.25$, and $0.5$, we have checked that the resultant dimers do not reside on the edges and the nature of edge states is not sensitive for changing the value of $Q_3'$. We therefore set $Q_3'/Q_3=0.5$ throughout this paper.

\section{Tomonaga-Luttinger liquid}\label{TLLiquid}
In order to judge whether or not a gapless edge mode is present, we utilize 
two-point spin-correlation functions at the edge of two models 
in Eq.~(\ref{eq:2models}). 
If it exists, an algebraic decay of the correlators is expected, while the correlators decay in an exponential fashion when the edge state has a 
finite excitation gap. 
When the dimerized model approaches the limit $\alpha\to\infty$, an isolated spin-$\frac{1}2$ AF chain appears at the edge and its low-energy physics 
is governed by a gapless Tomonaga-Luttinger (TL) liquid. 
It is therefore important to summarize spin corrrelators of the TL liquid phase as a reference point before embarking on our QMC results.

For an ideal TL liquid phase of spin-$\frac{1}{2}$ AF chains, 
transverse and longitudinal spin correlations are 
known to behave as~\cite{Giamarchi} 
\begin{subequations}
\label{eq:1dCorrelation}
\begin{eqnarray}
\label{eq:1dCorrelation1}
\langle S^{x}_{r}S^{x}_{0} \rangle & \sim & 
C_{1} (-1)^{r} r^{-\eta_{xy}}+\cdots,
\\
\langle S^{z}_{r}S^{z}_{0} \rangle & \sim &
m^{2} + C_{2} r^{-\eta_{z}}\cos(2 k_{F} r)+\cdots,
\label{eq:1dCorrelation2}
\end{eqnarray}
\end{subequations}
at long distances $r\gg 1$. Here the uniform magnetization 
$m=\langle S_j^z\rangle$ is induced by external magnetic field 
$H$ along the $S^z$ axis, 
$2k_{F}=\pi(1 - 2m)$ is the \"Fermi" wave number, and $C_{1,2}$ are 
non-universal constants. 
It is well known that critical exponents $\eta_{xy,z}$ satisfy the 
relation~\cite{Giamarchi}  
\begin{eqnarray}
\label{TLrelation}
\eta_{xy}\eta_z &= &1
\end{eqnarray}
and $\eta_{xy}=\eta_z=1$ occurs at the $SU(2)$-symmetric case. 
We also note that a $2k_F$ incommensurate oscillation disappears 
at $m=H=0$ in Eq.~(\ref{eq:1dCorrelation2}).

\section{Numerical Analysis}\label{Numerical}
This section is the main content of the paper, and we show 
all the important QMC results here. 
We discuss numerically evaluated  physical quantities, especially 
spin correlation functions of the two models~(\ref{eq:dimer}) and (\ref{eq:JQ3}),
and the stability of their edge states. 
In the QMC simulation, we adopt the boundary condition of 
Fig.~\ref{fig_model}, and mainly consider square-shaped finite-size systems 
in which the lengths of $x$ and $y$ directions, $L_x$ and $L_y$, are both 
fixed to $L$. In order to see the low-temperature physics of 
both models, we set temperature $T$ proportional to $L^{-1}$. 
We note that correlation lengths and critical exponents appear in this section 
are all evaluated from the QMC results for fixed models with 
the largest size and the lowest temperature.

\subsection{Gapless Edge States and Critical Exponents}\label{Gapless}
\begin{figure}[b]
\begin{center}
\includegraphics[width=7cm, trim=25 150 130 10]{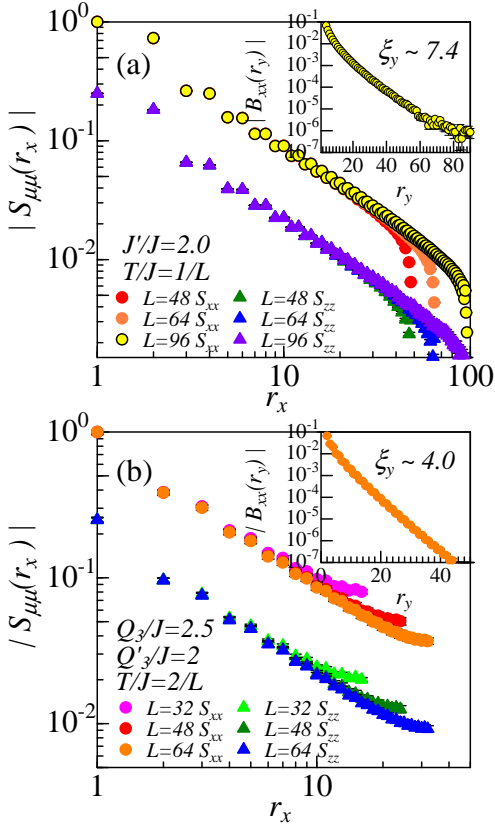}
\end{center}
\caption{(color online) 
Spin-spin correlation $|S_{\mu\mu}(r_x)|=
|\langle S^\mu_{(r_x,0)}S^\mu_{(0,0)}\rangle|$ on the edge 
in the dimerized phase of (a) dimerized model and (b) $JQ_3$ model. 
Insets are the results along the bulk direction: 
$|B_{\mu\mu}(r_y)|=|\langle S^\mu_{(L/2,r_y)}S^\mu_{(L/2,0)}\rangle|$. 
We normalize the values of $|S_{xx}(1)|$ and $|B_{xx}(1)|$ to be unity. 
In the inset, we depict only the results for $B_{xx}$ 
because of the $SU(2)$ symmetry of the models. 
Errors are drawn on all symbols, but they are less than the symbol size. }
\label{fig_zeroH}
\end{figure}
We first discuss the spin-correlation functions
in the dimer phases of the two models~(\ref{eq:dimer}) and (\ref{eq:JQ3}). 
Figure~\ref{fig_zeroH} shows numerically determined spin-correlation functions along the $x$ direction on edge and those along the $y$ direction at $x=L/2$ for the dimerized phases of both dimerized and $JQ_3$ models at sufficiently 
low temperatures $T\propto L^{-1}$. 
Hereafter, a correlation function along the $x$ direction on the edge (along the $y$ direction) is called edge (bulk) correlation. 
The long-distance behavior of edge spin correlations $S_{\mu\mu}(r_x)\equiv \langle S^\mu_{(r_x,0)}S^\mu_{(0,0)}\rangle$ is well explained by a power-law decay in both models, reflecting the existence of a gapless edge state. 
The critical exponent $\eta$ defined by $|S_{\mu\mu}(r)| \sim r^{-\eta}$ 
is evaluated as $\eta \simeq 1.13$ (0.97) for the dimer ($JQ_3$) model. 
We stress that {\it this algebraic-decay behavior survives far from 
the dimer limit}. 
In fact, the parameters $Q_3/J=2.5$ and $J'/J=2$ correspond to 
$Q_3/{Q_3}_c \simeq 1.67$ and $J'/J'_c \simeq 1.05$, respectively. 
Furthermore, another remarkable point is that 
{\it the critical exponent $\eta$ violates a TL liquid relation 
$\eta_{xy}=\eta_z=1$}. 
We have also evaluated $\eta \simeq 1.01$ ($0.99$) at a more deeply 
dimerized point $J'/J=5$ ($Q_3/J=4$) in the dimerized ($JQ_3$) model. 
On the other hand, as shown in the inset of Fig.~\ref{fig_zeroH}, the bulk spin correlations $B_{\mu\mu}(r_y)\equiv \langle S^\mu_{(L/2,r_y)}S^\mu_{(L/2,0)}\rangle$ decay exponentially, indicating a finite dimerization gap in the bulk. 
The correlation lengths $\xi_y$ of $|B_{\mu\mu}(r_y)| \sim \exp[-r_y/\xi_y]$ 
are evaluated as $\xi_y \sim 7.4$ (4.0) for the dimerized ($JQ_3$) model. 
These large values of $\xi_y$ clearly show that the edge chain is really 
correlated to the bulk. Namely, the edge correlation of Fig.~\ref{fig_zeroH} 
should be regarded as an intrinsic result of 2D models, and it should 
be distinguished from correlation functions in 1D systems such as 
spin chains and ladders. 
From these results, we conclude that a gapless edge state is realized in 
the wide range of both extrinsically and spontaneously spin-Peierls phases, 
but the critical exponent deviates from that of TL liquid, 
especially, near the 2D transition points between dimer and N\'eel phases.

In the rest of this subsection, let us consider 
why the critical exponent $\eta$ deviates from the 
value of the TL liquid $\eta_{xy}=\eta_z=1$. 
From the renormalization-group (RG) viewpoint,~\cite{Giamarchi,Sachdev} 
it would be natural that 
the value of $\eta$ approaches unity if the system is sufficiently close 
to the thermodynamic and zero-temperature limit. Such a tendency however 
cannot be observed upto the $L=96$ system in Fig.~\ref{fig_zeroH}. 
We have further checked that the exponent $\eta$ gradually decreases 
from 1.13, but it does not reach unity in the dimerized model 
at $J'/J=2$ as we increase both the edge length $L_x$ and 
the inverse of temperature $J/T=L_x$ upto 256 
by using rectangle-shaped systems with $L_x > L_y$.  
This result definitely suggests that very large sizes $L_{x,y}$ and 
extremely low temperatures are necessary to observe the crossover 
from the non-TL liquid to the usual TL liquid in the edge correlations 
if we approach the dimer-N\'eel transition point. In other words, 
the RG flow to the TL liquid fixed point is expected to be extremely slow. 
It is well known that a quantum critical region~\cite{Sachdev} 
is widely expanded around the dimer-N\'eel quantum critical 
point.~\cite{note_JQ3} The non-trivial value of $\eta$ is hence expected to 
be attributed to effects of large fluctuations around 
the 2D quantum critical point. 
In that sense, the values $\eta$ and their $J'/J$ and $Q_3/J$ dependence 
are characteristic properties of the 2D spin-Peierls systems, and 
they would not be observed in 1D quantum magnets such as $N$-leg spin ladders.

In experiments for spin-Peierls compounds, it is generally difficult 
to realize an extremely low temperature limit $T/J\to +0$ and 
a clean edge without any defect or any impurity. 
It is hence expected that the evaluated critical exponents $\eta\neq 1$ 
in large but finite systems can be relevant in real materials, in principle, 
rather than the value of the TL liquid $\eta_{xy,z}=1$. 
We will consider how to detect the gapless edge modes in the next section. 
Following the above argument, we will continuously use square-shaped systems 
with finite-size $L\times L$ to see the experimentally expected 
behavior of several correlation functions throughout this section.

\subsection{Stability against Perturbations}\label{Stability}
\begin{figure}[t]
\includegraphics[width=9cm, trim=5 500 0 0]{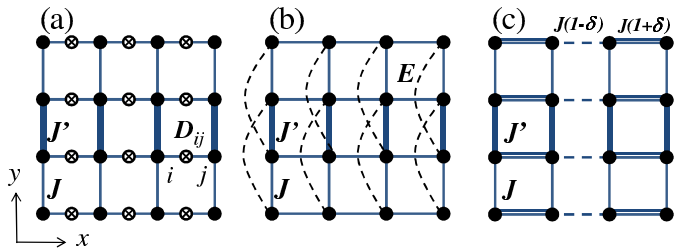}
\caption{(color online) Perturbations in the dimerized model: 
(a) DM interaction, (b) next-nearest neighbor interaction, and 
(c) bond alternation. 
Open boundary condition is imposed for both $x$ and $y$ directions, 
and the edge state appears parallel to the $x$ direction. 
In the panel (a), cross circles on $J$ bonds indicate 
the direction of ${\bol D}_{i,j}$ vector.}
\label{fig_model2}
\end{figure}

We next discuss how robust the gapless edge mode is against various kinds 
of perturbations. 
This is very important since the lack of stability indicates the 
difficulty of observing the edge state in real magnets. 
As realistic perturbations, we consider an uniform Zeeman term 
${\mathcal H}_{z}$, a XXZ-type magnetic anisotropy ${\mathcal H}_{\rm xxz}$, 
a Dzyaloshinskii-Moriya (DM) interaction ${\mathcal H}_{\rm DM}$ with 
DM vector $\bol D_{ij} \parallel S^z$, a next-nearest-neighbor interaction 
perpendicular to the edge direction ${\mathcal H}_{\rm N.N.N}$, and 
an additional bond modulation for the $x$ direction ${\mathcal H}_{\rm alt}$. 
They are expressed as
\begin{subequations}
\label{perturbations} 
\begin{eqnarray}
{\mathcal H}_{z} &=& -H \sum_j S^z_j,
\label{perturbation1}\\
{\mathcal H}_{\rm xxz} &=&-J\Delta\sum_{\langle i,j \rangle}S^z_{i} S^z_{j},
\label{perturbation2}\\
{\mathcal H}_{\rm DM}&=& 
\sum_{\langle i,j \rangle}  {\bol  D}_{ij} \cdot 
( {\bol S}_{i} \times {\bol S}_{j} ), \,\,\,\,\,(\bol D_{ij} \parallel S^z)
\label{perturbation3}\\
{\mathcal H}_{\rm N.N.N} &=& \sum_{i_x=j_x,i_y+2=j_y} 
E{\bol S}_i\cdot{\bol S}_j,
\label{perturbation4}\\
{\mathcal H}_{\rm alt} &=&J\sum_{r} (-1)^{r_{x}}
\delta {\bol S}_{r}{\bol S}_{r+{\bol e}_{x}},
\label{perturbation5}
\end{eqnarray}
\end{subequations}
and some of them are depicted in Fig.~\ref{fig_model2}. 
These perturbations possess the following nature of symmetry. 
The Zeeman term ${\mathcal H}_z$ breaks time-reversal symmetry, and 
${\mathcal H}_{\rm DM}$ breaks link-parity symmetry. 
These two and ${\mathcal H}_{\rm xxz}$ reduce the $SU(2)$ symmetry to 
the axial $U(1)$ type. 
The bond alternation term ${\mathcal H}_{\rm alt}$ eliminates the 
translational symmetry along the $x$ direction (if we apply a periodic 
boundary condition). 
In contrast, ${\mathcal H}_{\rm N.N.N}$ does not violate any symmetry 
of the original models.

\subsubsection{Effect of External Magnetic Field}
\begin{figure}[htb]
\begin{center}
\includegraphics[width=7cm, trim=25 265 200 10]{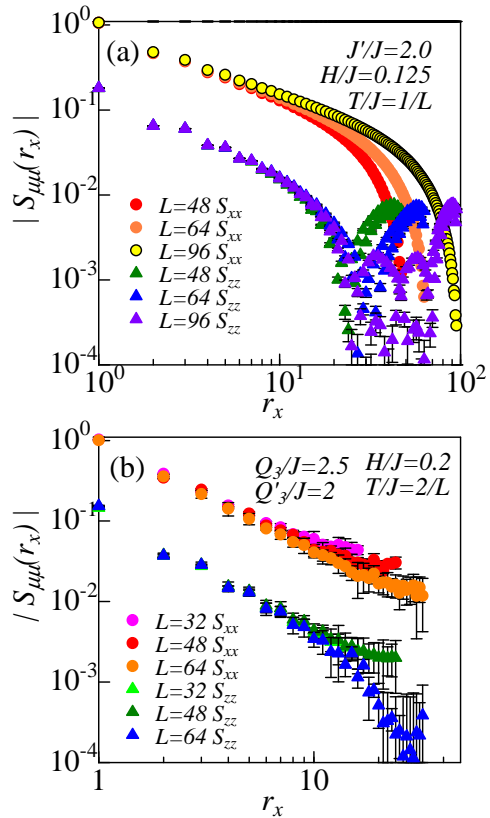}
\end{center}
\caption{(color online) 
Edge spin correlations $S_{\mu\mu}(r_x)$ of (a) dimerized model and 
(b) $JQ_3$ model under a magnetic field $H$. 
Circles and triangles are the transverse ($\mu=x$) and longitudinal 
($\mu=z$) correlators, respectively. We have eliminated a constant part 
of $\langle S_j^z\rangle^2$ from the data of $\mu=z$. 
Like Fig.~\ref{fig_zeroH}, we have used the normalization $S_{xx}(1)=1$.}
\label{fig_finiteH}
\end{figure}

\begin{figure}[htb]
\includegraphics[width=8cm, trim=0 0 0 0]{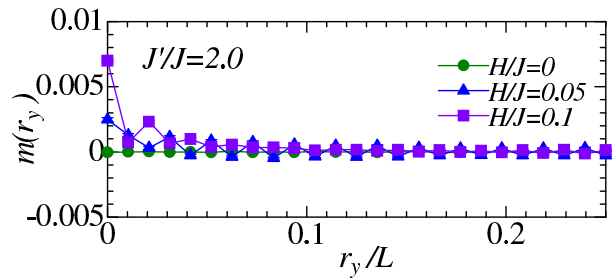}
\caption{(color online) Magnetization profile 
$m(r_y)=\langle S^z_{L/2,r_y}\rangle$ of the dimerized model with $J'/J=2$ 
under magnetic fields $H$ at $r_{x}=L/2$.}
\label{fig_magprofile}
\end{figure}

First let us consider the effects of the uniform magnetic field $H$ 
which is one of the few things we can control. We plot the edge spin 
correlations and magnetization profiles in the presence of magnetic field $H$ 
in Figs.~\ref{fig_finiteH} and \ref{fig_magprofile}, respectively. 
We have verified that the bulk correlation has a finite correlation length 
$\xi_y$ under the magnetic field $H$, i.e., the bulk is still gapped. 
Figure \ref{fig_finiteH} shows that even in $H$, a power-low decay fashion 
survives in the edge spin correlations, having an incommensurate 
oscillation in the longitudinal $S^z$ correlation. 
The stability against $H$ is in contrast to the helical edge of TIs and 
edge spins of Haldane states. 
The incommensurability is very similar to Eq.~(\ref{eq:1dCorrelation2}), 
while we have checked that such an incommensurate oscillation 
is absent in the bulk correlations. 
Figure~\ref{fig_magprofile} reveals that a finite magnetization emerges 
{\it only around the edge} with increase of $H$. 
Finite magnetizations on multiple sites near the edge indicate 
that in addition to the edge spin chain, some arrays around the edge 
also become gapless due to a small $H$. 
This inhomogeneous magnetization profile can be observed in principle, 
for example, by using nuclear magnetic resonance (NMR). 
Both Figs.~\ref{fig_finiteH} and \ref{fig_magprofile} clearly indicate 
that the gapless edge mode survives under field $H$. 

\subsubsection{Effects of Other Perturbations}
Next let us consider effects of 
${\mathcal H}_{\rm DM}$ and ${\mathcal H}_{\rm xxz}$. For the $JQ_3$ 
models with the perturbations 
(\ref{perturbation2})-(\ref{perturbation5}), 
technical difficulties of the QMC method emerge in highly accurate computations. 
We will therefore focus only on the dimerized model below. However effects of 
the perturbations on the dimerized model are probably very similar 
to those on the $JQ_3$ model since the wave functions of both extrinsically 
and spontaneously dimerized phases are expected to massively overlap 
each other.

Before the analysis, we should note the following property 
of the DM interaction. 
As we consider, for example, a uniform DM term with 
${\bol D}_{ij}=(0,0,D^{z})$ on the bonds along the $x$ direction 
[see Fig.~\ref{fig_model2}(a)], the Hamiltonian 
${\mathcal H}_{\rm dim}+{\mathcal H}_{\rm DM}$ can be mapped 
onto the following easy-plane anisotropic form
\begin{eqnarray}
\label{eq:modifiedDM}
{\mathcal H}'&=&\sum_{j} 
J_\perp(S^{x}_{j}S^{x}_{j+{\bol e}_x}+S^{y}_{j}S^{y}_{j+{\bol e}_x})
+JS^{z}_{j}S^{z}_{j+{\bol e}_x}\nonumber\\
&&+\sum_{j} J(1+\alpha\delta_{j_y,{\rm even}})
{\bol S}_{j}\cdot{\bol S}_{j+{\bol e}_y},
\end{eqnarray}
via a proper unitary transformation 
$S^\pm_{r} \rightarrow e^{i\theta_r}
S^\pm_{r}$.~\cite{DM_interaction1,DM_interaction2} 
Here $J_\perp=J\sqrt{1+(D^{z}/J)^{2}}$. 
Remarkably, the modified system~(\ref{eq:modifiedDM}) recovers the link-parity symmetry and has an easy-plane anisotropy. 
This kind of mapping can be applicable for a wide class of DM terms between neighboring spins if we adopt the open boundary condition for both $x$ and $y$ directions. 
Therefore, it is enough to study the dimerized model with an easy-plane XXZ anisotropy in order to see effects of both ${\mathcal H}_{\rm DM}$ and ${\mathcal H}_{\rm xxz}$.

\begin{figure}[htb]
\includegraphics[width=7cm, trim=20 10 160 0]{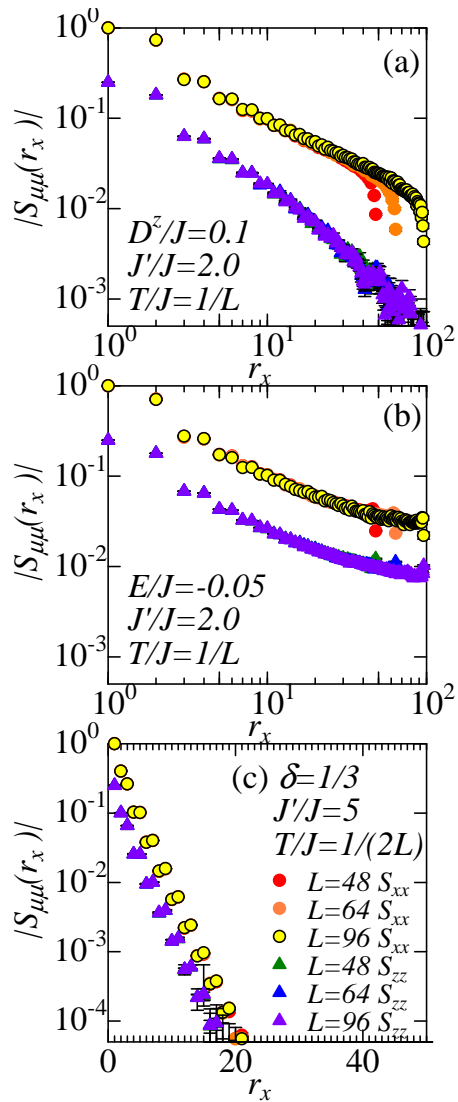}
\caption{\label{fig:6} (color online) 
Effects of (a) a DM interaction (equal to XXZ anisotropy), 
(b) a next-nearest neighbor interaction, and (c) an extra bond-alternation along the edge direction on the edge spin correlations of the dimerized model.
 Circles and triangles are the results of $\mu=x$ and $\mu=z$, respectively. 
In panel (c), we have used semi-log scale, and have set a larger $J'/J=5$ and a lower temperature $T=1/(2L)$ to depict a clear exponential-decay form of the edge correlation. In all of the panels, we have used the normalization 
$S_{xx}(1)=1$. }
\label{fig_perturb}
\end{figure}

Figure \ref{fig_perturb} (a) indicates that both the longitudinal and transverse edge spin correlations decay algebraically in the system (\ref{eq:modifiedDM}) with $D^z/J=0.1$. 
We thus conclude that the gapless edge state is stable against both the DM and easy-plane XXZ interactions. 
In purely 1D spin-$\frac{1}{2}$ AF chains with easy-plane anisotropy, critical exponents $\eta_{xy,z}$ usually satisfy $\eta_z>\eta_{xy}$ in addition to 
$\eta_z\eta_{xy}=1$.~\cite{Giamarchi} 
In the present 2D case of Fig.~\ref{fig_perturb} (a), however, we obtain 
$\eta_{xy} \simeq 0.92$ and $\eta_z \simeq 1.48$. 
Namely, the exponents satisfy the inequality $\eta_z>\eta_{xy}$, but 
the relation $\eta_z\eta_{xy}=1$ is clearly broken down. 
The difference between critical exponents of the 1D chain and the edge of 
the 2D dimerized model would be attributed to a strong correlation between 
bulk and edge.

The edge spin correlations for the case with ${\mathcal H}_{\rm N.N.N}$ are given in Fig.~\ref{fig_perturb} (b), in which we have adopted a ferromagnetic coupling $E/J=-0.05$ to avoid the negative sign problem. 
The figure shows a power-law decay of the correlation functions, and it indicates that the gapless edge mode still survives in a small 
${\mathcal H}_{\rm N.N.N}$. We, however, note that evaluated critical exponents 
$\eta_{xy} = \eta_{z} \sim 0.71$ deviate from the value of ideal TL liquid similarly to the cases with 
${\mathcal H}_{\rm DM}$ and ${\mathcal H}_{\rm xxz}$.

Figure \ref{fig_perturb} (c) is the result of the system 
${\mathcal H}_{\rm dim}+{\mathcal H}_{\rm alt}$. 
As expected, the edge spin correlation changes from an algebraic form to an exponential one due to the bond dimerization along the $x$ direction. 
From this analysis of perturbations, we see that the gapless nature of 
the edge state survives after introducing several perturbations with different 
symmetries except for the bond alternation ${\mathcal H}_{\rm alt}$. 
It suggests a high possibility of the realization of a gapless edge mode 
in 2D spin-Peierls compounds.

\section{How To Detect Edge Modes}\label{How}
Finally we consider possible experimental methods of probing signatures of 
the gapless edge modes in 2D spin-Peierls states. 
As we already mentioned, a finite magnetization rapidly grows only 
around the edge sites as shown in Fig.~\ref{fig_magprofile}, 
if we apply $H$ in the dimerized states. 
Observing such a site-dependent magnetization (e.g., by using NMR) 
could indicate a signature of the existence of a gapless edge state.

The NMR relaxation rate $1/T_1$ for nuclear spins near the edge is expected 
to contain a power-law $T$ dependence $T^{-\epsilon}$ 
at low temperatures,~\cite{Giamarchi} implying the existence 
of a gapless mode. If we assume that the dynamical critical 
exponent~\cite{Sachdev} $z$ of the gapless edge state is close enough 
to unity, similarly to the TL liquid, we can evaluate 
the temperature dependence of $1/T_1$ from a simple field-theory 
argument~\cite{Giamarchi} as follows:
\begin{eqnarray}
\label{eq:NMR_T1}
1/T_1 &=& A_{xy}T^{\eta_{xy}-1}+A_{z}T^{\eta_{z}-1}+\cdots,
\end{eqnarray}
where constants $A_{xy,z}$ depend on the strength and microscopic detail 
of interactions between electron and nuclear spins. 
This prediction of Eq.~(\ref{eq:NMR_T1}) indicates that critical exponents 
$\eta_{xy,z}$ of the edge mode can be measured from the NMR experiment 
in principle. 

\begin{figure}[htb]
 \begin{center}
   \includegraphics[width=6cm, trim=0 0 10 0, angle=270]{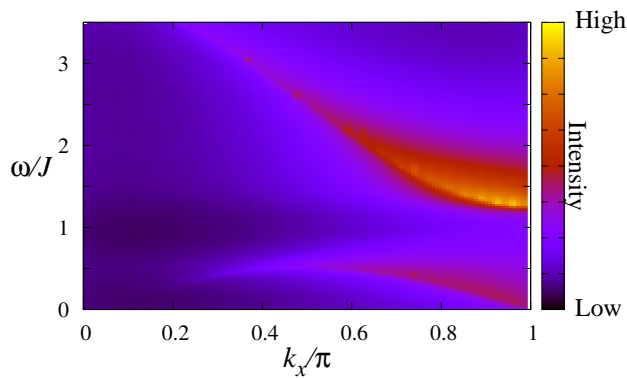}
 \end{center}  
\caption{\label{fig:7} (color online) 
Schematic inelastic-neutron-scattering spectrum of a 2D spin-Peierls phase 
with a gapless edge mode in the space of frequency $\omega$ and 
the wave number $k_x$. 
The wave number $k_{y}$ is fixed to $\pi$. 
The weight in the low-energy region $\omega \alt J$ is the contribution 
from the edge state, which lower bound would be similar to 
the des Cloizeaux-Peason mode $\sim J|\sin(k_{x})|$. 
The weight in the higher-energy region  $\omega \agt J$ comes from the 
gapped triplet excitations in the bulk.}
\label{fig_sqw}
\end{figure}

Inelastic neutron scattering spectra could also provide information 
about gapless edge modes. 
In the case of TIs, angular-resolved photoemission spectroscopy 
spectra~\cite{review1,review2} have been often used to see the gapless 
dispersion on the surface and edge in addition to gapped bulk excitations. 
Similarly, as shown in Fig.~\ref{fig_sqw}, the neutron-scattering spectra 
are expected to possess a des Cloizeaux-Peason-like gapless 
continuum~\cite{desCloizeauxPearson,Giamarchi} due to the edge state 
in 2D Peierls magnets. As we apply a small magnetic field $H$, 
the contribution from the longitudinal spin dynamics would have a peak 
at an incommensurate wave number $k_x\sim\pi(1-2 m_0)$ in which 
$m_0=m(r_y=0)$ is the magnetization on edge sites 
(see Fig.~\ref{fig_magprofile}). 
This gapless spectra would be strong evidence for the gapless edge mode. 
In addition to these ways, for instance, heat transport properties 
would capture the nature of the gapless edge mode.

\section{Conclusions}\label{Conclusion}
We have investigated several properties of the edge spins of 
the dimerized and $JQ_3$ models, Eqs.~(\ref{eq:dimer}) and (\ref{eq:JQ3}), 
by utilizing QMC simulation. The main numerical results are given 
in Sec.~\ref{Numerical}. 
When the system is in the dimerized phase and the singlet dimers do not 
reside on the edge, the gapless edge state really emerges. 
It is remarkable that the evaluated critical exponent $\eta$ of the edge spin 
correlation becomes different from the exponent $\eta=1$ of 
the usual TL liquid phases in large but finite systems at low temperatures. 
The deviation of $\eta$ becomes larger as we approach 
the dimer-N\'eel quantum phase transition point. 
This non-trivial value would be caused by the strong correlation 
between the edge and bulk with large quantum fluctuations. 
In the sense of RG, the edge exponent $\eta$ is naively expected 
to be reduced to the value of the
usual TL liquid at least in the thermodynamic and zero-temperature limit, 
but our QMC result strongly suggests that an extremely large system size 
and an extremely low temperature have to be prepared in order to observe 
such a crossover to $\eta\to 1$. Real experiments are usually done under 
low but finite temperatures and magnetic crystals generally contain 
impurities and some kinds of defects. 
Therefore, the non-trivial $\eta$ 
could be relevant and observed in real spin-Peierls materials rather 
than the ideal asymptotic value $\eta=1$ of the 1D AF spin chain. 
The non-trivial value of $\eta$ and its change depending on 
the ``distance" from the quantum transition point are unique features of 
the edge state in 2D spin-Peierls phases, and they do not 
appear in well-studied 1D magnets such as spin ladders.

We have also shown that the edge mode is quite robust against 
various perturbations with different symmetries in Sec.~\ref{Stability}. 
Particularly, the stability against external magnetic field $H$ is 
in contrast with the helical edge modes of TIs and free edge spins of 
the Haldane state, and $H$ induces an inhomogeneous magnetization around the 
edge as shown in Fig.~\ref{fig_magprofile}. 
The edge spin correlations algebraically decay like a TL liquid, 
but their critical exponents $\eta_{xy,z}$ generally violate the TL liquid 
relation $\eta_{xy}\eta_z=1$, depending on the detail of the models. 
As expected, the deviation from the TL liquid becomes larger 
when coupling constants of perturbations are stronger.

We finally consider some experimental methods of detecting 
signatures of gapless edge modes in Sec.~\ref{How}. 
NMR, inelastic neutron scattering, and heat transport would provide 
hopeful experimental ways. In particular, the NMR relaxation rate $1/T_1$ has 
the potential to measure the value of non-trivial critical exponents 
$\eta_{xy,z}$ of the edge correlation.

Properties of gapless edge states we illustrate in this paper 
would also be useful as we study other gapless edge states of 
more exotic quantum spin systems. 
Recently, some kinds of topologically ordered states~\cite{XGWen,Misguich} 
such as $\mathbb{Z}_2$ spin liquids 
have been predicted in relatively realistic quantum spin models, 
especially, frustrated models. 
The realization of such exotic states is generally difficult, 
but the comparison between their edge states and that of spin-Peierls magnets 
would be an interesting direction of theoretical studies in order 
to characterize the exotic edge states.

\section*{Acknowledgment}
We thank J. Lou and H. Tsunetsugu for fruitful discussions. 
Numerical calculations were performed at the Institute for Solid State Physics Supercomputer Center of the University of Tokyo and cluster machines in Nano-Micro Structure Science and Engineering, University of Hyogo.


\begin{thebibliography}{99}
\bibitem{note_JQ3}
The dimer-N\'eel transition of the dimerized model is well believed to 
be a second-order type quantum transition. On the other hand, 
we should note that it is still controversial whether 
the dimer-N\'eel transition of the $JQ_3$ model 
is of a continuous type or a weak first-order type.
\end{thebibliography}

\end{document}